\documentclass[referee]{raa}            

\usepackage{graphicx,times}             
\usepackage{natbib}
\usepackage{amssymb,amsmath}
\bibpunct{(}{)}{;}{a}{}{,}

\usepackage[a4paper=true,dvipdfm=true,pagebackref=true]{hyperref}
\hypersetup{colorlinks = true, linkcolor = green, anchorcolor = red, citecolor = blue, filecolor = red, pagecolor = red, urlcolor = red}

\usepackage{graphicx}
\usepackage{color}
\newcommand{\Ha}{\mbox{H$\alpha$}}      
\newcommand{\HII}{\ion{H}{II}}    

\begin{document}

   \title{Spatially-resolved Stellar Population Properties of the M 51--NGC 5195 System from Multi-wavelength Photometric Data}

   \volnopage{Vol.0 (20xx) No.0, 000--000}      
   \setcounter{page}{1}          

   \author{Peng Wei
      \inst{1,2,3,4}
   \and Hu Zou
      \inst{3,4}
   \and Lin Lin
      \inst{5}
   \and Xu Zhou
      \inst{3,4}
   \and Xiang Liu
      \inst{1}
   \and Xu Kong
      \inst{6}
   \and Lu Ma
      \inst{1}
   \and Shu-Guo Ma
      \inst{1,2}
   }

   \institute{Xinjiang Astronomical Observatory, Chinese Academy of Sciences, Urumqi 830011, P. R. China; {\it weipeng@xao.ac.cn}\\
        \and
             School of Astronomy and Space Science, University of Chinese Academy of Sciences, Beijing 101408, P. R. China\\
        \and
             CAS Key Laboratory of Optical Astronomy, National Astronomical Observatories, Chinese Academy of Sciences, Beijing 100101, P. R. China; {\it zouhu@nao.cas.cn}\\
        \and
             Center for Astronomical Mega-Science, Chinese Academy of Sciences, 20A Datun Road, Chaoyang District, Beijing, 100012, P. R. China\\
        \and
             Shanghai Astronomical Observatory, Chinese Academy of Sciences, Shanghai 200030, P. R. China\\
        \and
             Key Laboratory for Research in Galaxies and Cosmology, Department of Astronomy, University of Science and Technology of China, Hefei 230026, P. R. China\\
\vs\no
   {\small Received~~20xx month day; accepted~~20xx~~month day}
   }

\abstract{Using multi-band photometric images of M 51 and its companion NGC 5195 from ultraviolet to optical and infrared, we investigate spatially resolved stellar population properties of this interacting system with stellar population synthesis models. The observed IRX is used to constrain dust extinction. Stellar mass is also inferred from the model fitting. By fitting observed spectral energy distributions (SEDs) with synthetical ones, we derive two-dimensional distributions of stellar age, metallicity, dust extinction, and stellar mass. In M51, two grand-designed spiral arms extending from the bulge show young age, rich metallicity, and abundant dust. The inter-arm regions are filled with older, metal-poorer, and less dusty stellar populations. Except for the spiral arm extending from M 51 into NGC 5195, the stellar population properties of NGC 5195 are quite featureless. NGC 5195 is much older than M 51, and its core is very dusty with $A_V$ up to 1.67 mag and dense in stellar mass surface density. The close encounters might drive the dust in the spiral arm of M51 into the center of NGC 5195. 
}

 \authorrunning{P. Wei et al.}            
 \titlerunning{M 51--NGC 5195 System }  

\maketitle

%
%
\section{Introduction} \label{introduction}   

Understanding how galaxies have evolved into their present-day forms is one of the foremost goals in extragalactic astronomy. At present, the evolution of local galaxies might be in transition from hierarchical clustering and merging in the early universe to secular processes in the future \citep{Kormendy04}. Hierarchical clustering and merging processes are rapid and violent and can account remarkably well for large-scale structures in the cosmological $\Lambda$ cold dark matter ($\Lambda$-CDM) models \citep{white78}.  For nearby galaxies, there have been a considerable number of studies on the galaxy formation and evolution induced by some secular processes, gas-rich accretion events, gravitational encounters, and gravitational interactions with satellite galaxies \citep{toomre72, Kormendy04, cor06, fis08, roskar08, fisher09, fis10, Kormendy10}.  

Nearby galaxies provide ideal astrophysical laboratories to study galaxy formation and evolution in sub-galactic scales, both observationally and theoretically \citep{dob10}. There are substantial multi-band archival data from various surveys and telescopes, such as the Galaxy Evolution Explorer \citep[GALEX;][]{martin05}, Sloan Digital Sky Survey \citep[SDSS;][]{york00}, Beijing-Arizona-Taiwan-Connecticut (BATC) Multicolor Sky Survey \citep{bur94}, Two Micron All Sky Survey \citep[2MASS;][]{jarrett03}, Spitzer Space Telescope \citep{Werner04,kennicutt03}, Herschel Very Nearby Galaxies Survey \citep[VNGS;][]{men12}, and The HI Nearby Galaxy Survey \citep[THINGS;][]{walter08}. Multi-wavelength data from ultraviolet to radio, which contain abundant information of stellar, gas and dust contents, can help us to understand the galaxy evolution.

As evolutionary population synthesis (EPS) models get more and more popular and mature \citep{fio97,leitherer99,bru03,kotulla09}, multi-band photometric data have been used to accurately derive physical properties of underlying stellar populations, such as stellar age, metallicity, and dust extinction. Based on the assumption that stars in a small region formed in an instantaneous burst that can be modeled as a simple stellar population (SSP), \citet{kong00} made use of 13 BATC intermediate-band images of M 81 and SSP models of GISSEL96 \citep{cha91,bru93} to derive two-dimensional age and reddening distributions. In this study, one of BATC near-infrared color was used to estimate the metallicity. However, galaxies are much more complex systems than SSPs. Composite stellar populations (CSPs) assuming a proper star formation history should be more accurate to model observational features of galaxies. \citet{li04b,li04a} utilized the PEGASE CSP models \citep{fio97} with exponentially decreasing star formation rate (SFR)  to analyze the stellar population properties of M 81 and M 33. They obtained reasonable age maps of these two galaxies based on images of a broad $U$ band and 13 intermediate bands from the BATC survey. \citet{zou2011} and \citet{lin13} utilized the UV-to-IR SEDs and BC03 models \citep{bru03} to analyze spatially resolved stellar population properties of NGC 628 and M 101. Furthermore, the empirical IRX-A$_\mathrm{FUV}$ relation is used to constrain the dust extinction in \citet{lin13}. In these studies, they derived reliable two-dimensional maps of age, metallicity, dust reddening, and corresponding radial profiles. Both NGC 628 and M 101 were reported to have features of pseudobulges and secular evolution. 

In our previous works, we have analyzed the detailed two-dimensional stellar population properties for several relatively isolated galaxies. This paper aims to study an interacting galaxy pair of M 51 and NGC 5195. M 51 (NGC 5194, the Whirlpool nebula) is a grand-design face-on spiral galaxy with Hubble type of Sbc. It has an inclination angle of about $20^\circ$ and a distance of about 8.4 Mpc \citep{fel97}. NGC 5195 is a post-starburst galaxy, which has no recent star formation \citep{men12}. It was found that these two galaxies underwent a close encounter about 300--500 Myr ago by kinematical and hydrodynamical simulations \citep{salo00,dob10}. Through stellar population synthesis modeling, \citet{men12} also found that M 51 and NGC 5195 underwent a burst of star formation roughly 370--480 Myr ago, which is consistent with the simulations. In their study, they focused on the dust properties derived from mid- and far-infrared data and meanwhile obtained the spatially-resolved distributions of stellar population properties, such as age, metallicity, dust extinction, and star formation time scale. However, the determination of stellar population properties in \citet{men12} was based on only 7 bands, including 4 optical bands and three near infrared bands. Besides, two components of star formation histories were adopted. Therefore, the parameter degeneracy might be relatively serious. For example, they claimed that their dust extinction was modestly underestimated.  

The M 51-NGC 5195 system has been observed by the BATC 15 intermediate-band filters. The intermediate color can be used to well constrain the metallicity \citep{kong00}. In addition, there are abundant archival multi-wavelength photometric data ranging from ultraviolet to infrared. A total of 26 bands are gathered with wavelength ranging from 1500 {\AA} to 4.5 $\mu$m. These bands are dominated by star light and are used to derive the stellar population properties. The wide wavelength coverage and the dust constraint by adopting the IRX as used in \citet{lin13} help to degrade the parameter degeneracy significantly. We can also get higher spatial-resolution maps of those properties (about 6{\arcsec} or 240 pc, while $\sim$1 kpc in \citet{men12}) and thus study the interacting system in more details. Through the derived distributions of the stellar population properties, we try to probe evolutionary clues of the M 51-NGC 5195 system and investigate possible influence of the galactic interaction.

 The outline of this paper is as follows. Section \ref{sec:data} introduces the multi-wavelength data and related image processing. Section \ref{sec:method} describes the stellar population synthesis models and corresponding SED fitting method. The parameter uncertainty and test of the parameter degeneracy are analyzed in this section. Section \ref{sec:result} presents the distributions of stellar population properties and Section \ref{sec:discussion}   shows some discussions. Section \ref{sec:summary} gives a summary.


\section{Multi-wavelength data and image processing} \label{sec:data}

Multi-wavelength photometric images of M 51 and NGC 5195 are collected from a series of surveys and telescopes. A total of 28 images are used in this paper, including 2 UV bands from $Galaxy ~Evolution ~Explorer$, 1 UV band from $XMM~Optical~Monitor$, 15 BATC intermediate bands, 3 broad bands from Beijing-Arizona Sky Survey \citep[BASS;][]{zou2017} and Mayall z-band Legacy Survey \citep[MzLS;][]{silva16}, 3 near-IR bands from 2MASS, and 4 near- and mid-IR bands from Spitzer. GALEX FUV and Spitzer mid-IR images are used to calculate dust extinction. The rest images are used to derive the stellar population properties. Table \ref{tab:obs} summarizes the information of these data.

\begin{table*}
\centering
\caption{Information for multi-band data of M 51-NGC 5195 galaxy pair} 
\label{tab:obs}
\begin{tabular}{cccccccc}
\hline
Name$^a$ & Filter & $\lambda_{\rm eff}$$^b$ & Bandwidth$^c$ & Pixel scale$^d$ & FWHM$^e$ &  Calibration$^f$ & Reference$^g$ \\
\hline
GALEX & FUV & 1516 & 268 & 1.5 & 4.85 & 0.05 & (1) \\
      & NUV & 2267 & 732 & 1.5 & 5.52 & 0.03 &     \\
\hline
XMM-OM & UVW1 & 2905 & 620  & 0.95 & 2.67 & 0.03 & (2) \\
\hline
BATC & $a$-$p$ & 3000-9900 & 120-310 & 1.7 & 3.2-4.8 & 0.03 & (3) \\
\hline
BASS & $g$ & 4776 & 848 & 0.45 & 1.92 & 0.01 & (4) \\
     & $r$ & 6412 & 833 &  0.45 & 1.70 & 0.01 &    \\
MzLS & $z$ & 9203 & 826 & 0.26 & 1.20 & 0.01 & (5) \\
\hline
2MASS & $J$ & 12350 & 1620 & 1.0 & 3.31 & 0.03 & (6) \\
      & $H$ & 16620 & 2510 & 1.0 & 3.25 & 0.03 &     \\
      & $K_{s}$ & 21590 & 2620 & 1.0 & 3.35 & 0.03 &  \\
\hline
$Spitzer$ & IRAC1 & 35500 & 7500 & 0.75 & 2.25 & 0.03 & (7) \\
        & IRAC2 & 44930 & 10100 & 0.75 & 2.27 & 0.03 &      \\
        & IRAC4 & 78720 & 29300 & 0.75 & 2.65 & 0.03 &       \\
        & MIPS24 & 237000 & 47000 & 2.5 & 6.0 & 0.03 &      \\
\hline

\multicolumn{8}{l}{$^a$ Name of the survey or telescope.}\\
\multicolumn{8}{l}{$^b$ Effective wavelength of each filter in \AA.}\\
\multicolumn{8}{l}{$^c$ Bandwidth of each filter in \AA.}\\
\multicolumn{8}{l}{$^d$ Pixel scale in arcsec.}\\
\multicolumn{8}{l}{$^e$ FWHM of stars in arcsec. }\\
\multicolumn{8}{l}{$^f$ Roughly estimated flux calibration accuracy in mag. }\\
\multicolumn{8}{l}{$^g$ References: (1)\citet{morrissey07}; (2)\citet{kuntz08}; (3)\citet{zhou01}; (4)\citet{zou2017,zou2017aj,zou18}; }\\
\multicolumn{8}{l}{ (5)\citet{silva16}; (6)\citet{skrutskie06}; (7)\citet{kennicutt03}.}\\
\end{tabular}
\end{table*}

\subsection{Ultraviolet images} \label{sec:GALEX}
Ultraviolet images come from GALEX \citep{martin05} and XMM-OM \citep{mas01}. GALEX is a 50 cm Ritchey-Chretien telescope and has two simultaneous channels of Far-UV (FUV) and near-UV (NUV) with effective wavelengths at 1516 and 2267 \AA, respectively.  Deep images of M 51 in these two bands were taken in 2007 with a total exposure time of 10,787 seconds by the Guest Investigator Program. These data are retrieved from GALEX GR6/GR7 data release\footnote{http://galex.stsci.edu/GR6/}. The XMM-OM is an optical/UV 30 cm telescope co-aligned with X-ray telescopes. It has a field of view of about 17{\arcmin} and covers a wavelength range of 1600--6000 \AA. The telescope has three UV filters, but M51 was observed in only UVW1 band. The UVW1 mosaic can be obtained at MAST \footnote{\url{http://archive.stsci.edu/index.html}}.

\subsection{BATC Intermediate-band images} \label{subsec:optical}
M 51 was observed by the 60/90 cm Schmidt telescope at the Xinglong Station of National Astronomical Observatory of China, as part of the BATC sky survey. A description of this survey and the observing strategy can be found in \citet{bur94}. A $2048\times2048$ Ford Aerospace CCD camera with a pixel scale of $1\arcsec.7$ was mounted at the focal plane of the telescope. The field of view is about $58\arcmin$. The photometric system consists of 15 dedicated intermediate-band filters covering the wavelength range of 3300--10000 {\AA} with bandwidths of about 200--300 \AA. These filters are well designed to avoid strong sky emission lines \citep{fan96}. The observations of the M 51 field started in 1995 January and ended in 2007 July. Normally, the individual exposure time was about 20 minutes. The raw images are processed by a data reduction pipeline customized for the BATC survey. The pipeline generates calibrated images with astrometric and photometric solutions tied to the UCAC3 catalog \citep{zacharias10} and four Oke-Gunn standard stars \citep{oke83}, respectively. The standard stars were observed on photometric nights, giving a typical calibration accuracy of about 3\%. The single-epoch images are aligned and stacked to create deep mosaics. Table \ref{tab:batcobs} lists some observational statistics and filter parameters.

\begin{table}
\centering
\caption{Filter Parameters for the BATC photometric system and observational statistics of the M 51 field}
\label{tab:batcobs}
\begin{tabular}{cccccc}
\hline
No. & Filter & $\lambda_{\rm eff}$$^a$ & Bandwidth & Exposure$^b$ & FWHM$^c$ \\
\hline
1 & a & 3360 & 222 & 1080 & 4.88 \\
2 & b & 3890 & 187 & 25200 & 4.45 \\
3 & c & 4210 & 185 & 37500 & 4.03 \\
4 & d & 4550 & 222 & 19200 & 4.17 \\
5 & e & 4920 & 225 & 13200 & 4.09 \\
6 & f & 5270 & 211 & 26400 & 4.13 \\
7 & g & 5795 & 176 & 8400 & 3.43 \\
8 & h & 6075 & 190 & 3600 & 4.53 \\
9 & i & 6660 & 312 & 3600 & 4.25 \\
10 & j & 7050 & 121 & 2700 & 3.99 \\
11 & k & 7490 & 125 & 8400 & 3.81 \\
12 & m & 8020 & 179 & 7200 & 4.00 \\
13 & n & 8480 & 152 & 13200 & 3.63 \\
14 & o & 9190 & 194 & 10800 & 3.37 \\
15 & p & 9745 & 188 & 3600 & 4.82 \\
\hline
\multicolumn{6}{l}{$^a$ Effective wavelength in \AA.} \\
\multicolumn{6}{l}{$^b$ Total exposure time in seconds.} \\
\multicolumn{6}{l}{$^c$ FWHM of stars in arcsecs.} \\
\end{tabular}
\end{table}

\subsection{Optical broad-band images} \label{sec:BASS and MzLS}
M 51/NGC5195 is also observed by BASS and MzLS. These two surveys serve for the spectroscopic targeting of the Dark Energy Spectroscopic Instrument \citep{dey18}.  The BASS uses a wide-field camera of 90Prime deployed on the 2.3 m Bok telescope at Kitt Peak. The MzLS uses the MOSAIC-3 camera of the 4 m Mayall telescope on the same mountain. The photometric filters include $g$ and $r$ bands for BASS and $z$ band for MzLS. These filters are very close to the ones used by the Dark Energy Survey \citep[DES;][]{des16}. There are three individual exposures for each filter. These individual CCD frames are reduced by the BASS pipeline \citep{zou2017aj,zou18}. The astrometric and photometric calibrations are respectively tied to the Gaia DR1 \citep{gai16} and Pan-STARRS1 catalogs \citep{cha16}. We stack individual images into deep mosaics by SWarp \citep{bertin02}. The mosaics have a size of 5400$\times$5400 pixels with a pixel scale is 0\arcsec.27, which is close to the MOSAIC-3 CCD pixel size. The image qualities in PSF FWHM as shown in Table \ref{tab:obs} are 2\arcsec.0, 1\arcsec.8, and 1\arcsec.3 for $g$, $r$, and $z$ bands, respectively.

\subsection{Near/mid-infrared images} \label{sec:2MASS&Spizter}
The near-infrared images of $JHK_s$ bands with central wavelengths of 1.2, 1.65, and 2.2 $\mu$m are from the 2MASS Large Galaxy Atlas survey \citep{jarrett03}. The PSF FWHMs of 2MASS images are estimated to be about $3\arcsec.3$. Other infrared observations come from the Spitzer Infrared Nearby Galaxies Survey \citep[SINGS;][]{kennicutt03}. M 51 was imaged by Spitzer with both Infrared Array Camera (IRAC at 3.6, 4.5, 5.6, $8.0\mu$m) and Multi-band Imaging Photometer (MIPS at 24, 70, and $160\mu$m). The PSF FWHM for IRAC is about 2\arcsec.0 and that for MIPS is larger than 6\arcsec.

\subsection{Image Processing} \label{sec:image processing}
Our goal is to analyze spatially resolved stellar population properties, so we need to uniformize the images from different telescopes (e.g., pixel scale and resolution). Here we describe the processing steps shortly \citep[see more details in][]{zou2011,lin13}. Firstly, the sky background is determined by a dedicated algorithm and subtracted from the mosaic for each band. The background map is a 2D polynomial fitting of the background pixels after signals from objects and central large galaxies are removed. Secondly, all images are convolved to the MIPS 24 $\mu$m resolution (FWHM of about 6\arcsec) with kernels estimated from the PSF profiles. Thirdly, all images are projected to the BATC $i$ band with a pixel scale of $1\arcsec.7$. Fourthly, foreground bright stars identified from 2MASS Point Source Catalog \citep{skrutskie06} are masked. 

\begin{figure*}
\centering
 \includegraphics[width=0.9\textwidth,angle=-90]{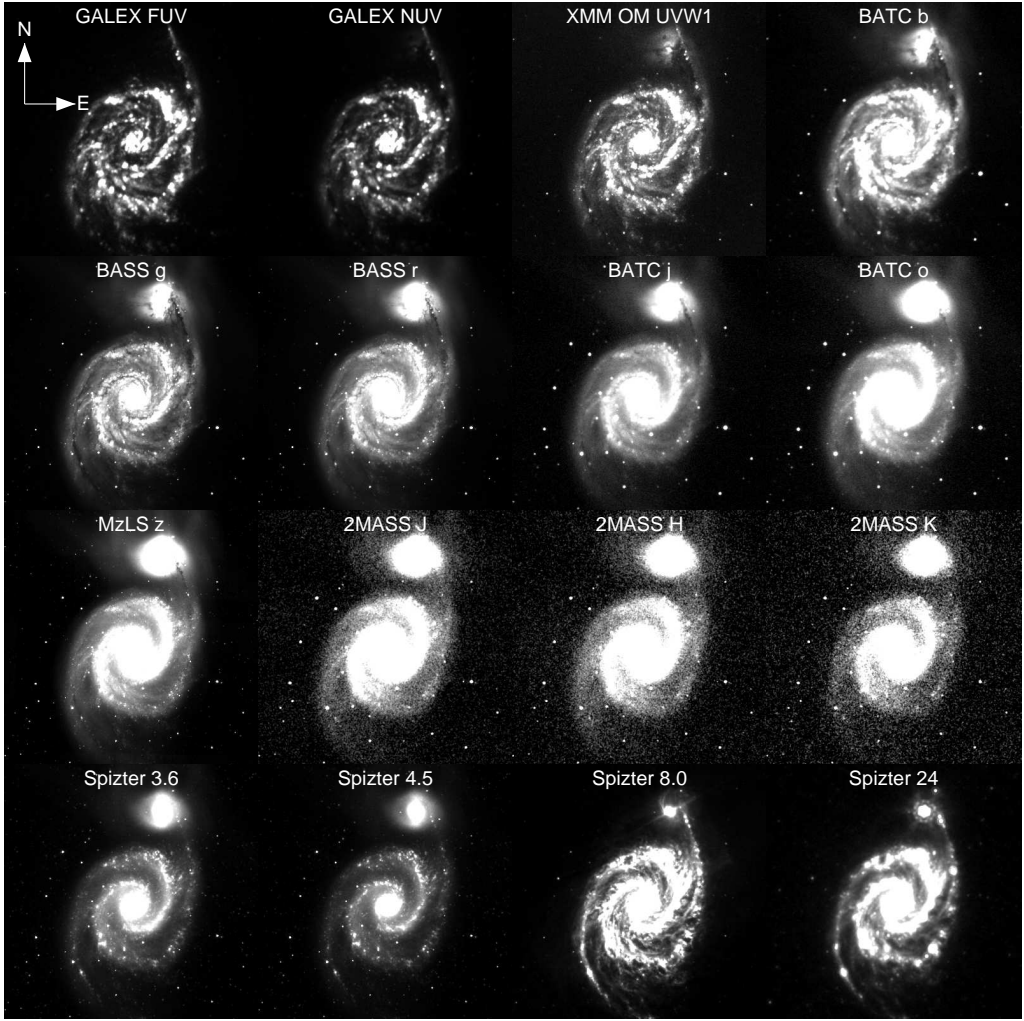}
 \caption{All processed images from GALEX, XMM-OM, BATC, BASS, MzLS, 2MASS, and Spitzer. The image size is about $11\arcmin\times11\arcmin$, corresponding to $27\times27$ kpc at a distance of 8.4 Mpc. The pixel scale is $1.\arcsec7$. North is up and east is right.}
\label{fig:allfit}
\end{figure*}

Figure \ref{fig:allfit} shows processed multi-wavelength images of the M 51--NGC 5195 pair from different telescopes. Ultraviolet and mid-infrared emissions trace the star formation and dust distribution. Near infrared bands are less affected by  dust attenuation and can trace the stellar mass. From these images, the SED is extracted pixel by pixel. It is corrected for the Galactic extinction using the extinction law of \citet{car89} and  the reddening map of \citet{schlegel98} ($E(B-V)$ = 0.031).

\section{Stellar population models and SED fitting method} \label{sec:method}
\subsection{Spectral models}\label{sec:models}
The evolutionary population synthesis becomes a modern popular technique that can be used for studying the stellar population properties of star clusters and galaxies. We can obtain a series of physical properties such as age, metallicity, intrinsic reddening, redshift, stellar mass, and star formation rate by comparing observed spectra or photometric SEDs with model ones. Some popular stellar population synthesis models include PEGASE \citep{fio97}, StarBurst99 \citep{leitherer99}, GALEXEV \citep[][hereafter BC03]{bru03}, Ma05 \citep{maraston05}, CB07 \citep{bru07}, BPASS\citep{eld09}, and GALEV \citep{kotulla09}, etc.

The simplest EPS model is the simple stellar population (SSP), which is composed of coeval stars with the same age and chemical composition. Providing an initial mass function (IMF) and stellar evolutionary tracks as well as stellar libraries, one can obtain the spectral evolution of an SSP.  SSPs are suitable for star clusters and {\HII} regions. However, galaxies are much more complex systems. An evolutionary composite stellar population (CSP) can more accurately describe the evolution history of a galaxy. The CSP is considered to be a superimposition of SSPs at different ages through assuming a star formation history (SFH). 

In this paper, we use a popular stellar population synthetical code of BC03. The adopted parameter configurations include the \citet{cha03} IMF, Padova 1994 evolutionary tracks, and delayed-exponential SFH. The delayed-exponential SFH is expressed as $\psi(t) = \frac{t}{\tau^{2}}\mathrm{exp}\left(\frac{-t}{\tau}\right)$, where $t$ is age and $\tau$ is the star formation timescale. As described in \citet{lin13}, we construct a library of 10$^{5}$ CSP model spectra, which are randomly sampled in different parameter spaces of age, metallicity, and $\tau$. The age is uniformly sampled between 1 Myr and 13.5 Gyr. The metallicity is uniformly sampled between 0.2 and 2 $Z_{\sun}$. The star formation timescale $\tau$ is sampled according to the probability density function of $p(\tau) = 1 - \mathrm{tanh}\left(\frac{8}{\tau}-6\right)$ with $\tau^{-1}$ ranging from 0 to 1.

\subsection{SED fitting method and constraint of dust extinction}\label{sec:fitting}
With the spectral library of CSP models as created above, we can construct multi-band model SEDs by convolving the reddened model spectra with filter responses if the intrinsic dust extinction $A_V$ is specified. A $\chi^2$ minimization is performed to estimate the stellar population parameters:  
\begin{equation}
\chi^2=\sum\limits_{i=1}^{n}{\frac{[F^\mathrm{obs}_{ i}-c{\times}F^\mathrm{csp}_{i}(t,Z,\tau, A_V)]^2}{\sigma_{i}^2}},
\end{equation}
where $n$ is the number of filters used for fitting, $F^\mathrm{obs}_{i}$ and $\sigma_{i}$ represent the observed flux and corresponding error for the $i$th filter, $c$ is a scaling factor, and $F^\mathrm{csp}_{i}$ is the integrated CSP model flux at specified age $t$, metallicity $Z$, star formation timescale $\tau$, and dust extinction $A_V$ for the $i$th filter. The age, metallicity, and $\tau$ are taken as free parameters and $A_V$ is constrained by other observational quantities.

As presented in \citet{lin13}, the dust extinction can be constrained using combined infrared and ultraviolet observations in order to degrade the parameter degeneracy. The dust attenuation can be estimated by IRX, which is the ratio of total IR (TIR) and FUV luminosities \citep{meurer99,hao11}. Here, we use observed IRX to constrain the dust extinction. First, we redden each CSP spectrum for a randomly given $A_V$ (ranging from 0 to 3 mag), assuming the extinction law of \citet{car89}. Second, the synthetic IRX is calculated through the energy conservation, where the total IR luminosity is the flux loss before and after reddening and FUV luminosity is computed by convolving the reddened model with the filter response. Our SED fitting is limited to those models whose synthetic IRXs are close to the observed IRX within 0.1, considering the measurement uncertainties. The observed IRX is calculated from IRAC 8 $\mu$m and MIPS 24 $\mu$m luminosities \citep{cal05}, which is formulated as
\begin{equation}
\mathrm{IRX} = \log{\frac{L_\mathrm{TIR}}{L_\mathrm{FUV}}}=\log{L(24)}+0.793\log{\frac{L_\nu(8)}{L_\nu(24)}}+0.908-\log{L_\mathrm{FUV}},
\end{equation}
where $L_\mathrm{TIR}$, $L_\mathrm{FUV}$, $L(24)$ is the total infrared, FUV, 24 $\mu$m luminosities and $L(24)=\nu L_\nu(24)$. 

\subsection{Error estimation and parameter degeneracy} \label{subsec:error}
\begin{figure*}
\centering
\includegraphics[width=0.65\textwidth,angle=-90]{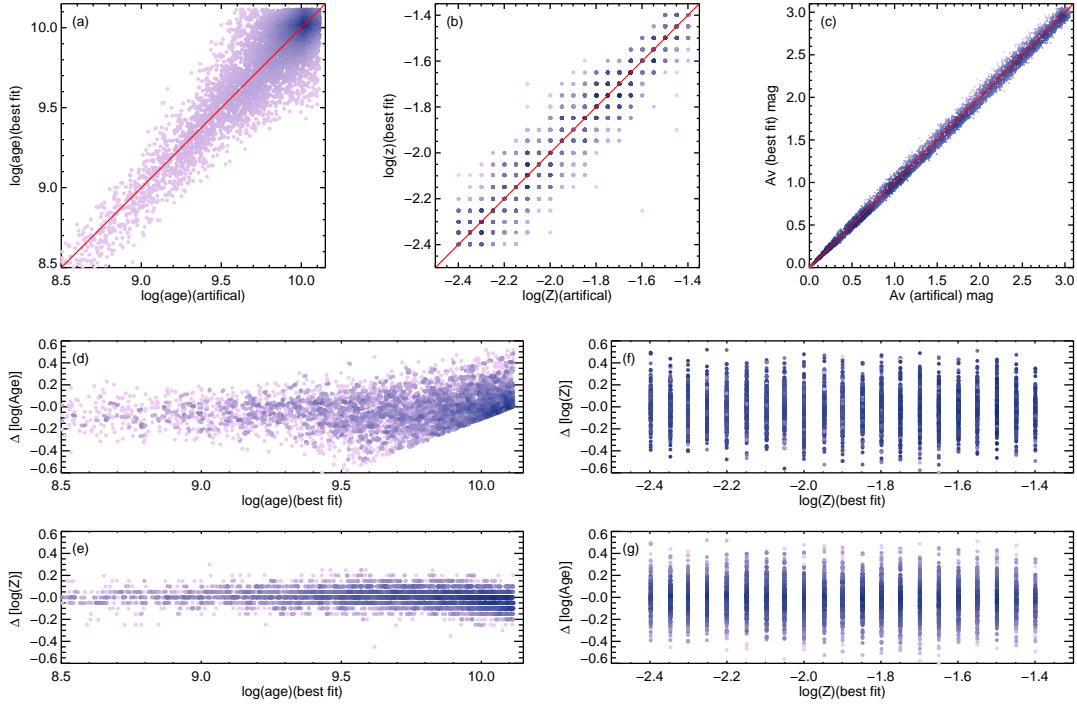}
\caption{Comparisons of the best-fitted parameters determined by our SED-fitting method and artificial parameters based on the simulation at a typical S/N of 20. (a) is for age, (b) is for metallicity, and (c) is for $A_V$. The units are yr for age, dex for $\log(Z)$, and mag for $A_V$. The diagonal solid line in each panel shows the equality between the fitted and artificial parameters. (d) $\Delta \log \mathrm{Age}$ as a function of $\log \mathrm{Age}$. (e) $\Delta \log Z$ as a function of $Z$. (f) $\Delta \log Z$ as a function of $\log \mathrm{Age}$. (g) $\Delta \log \mathrm{Age}$ as a function of $\log Z$. $\Delta \log Z$ is the difference between artificial metallicity and best-fitted value and $\Delta \log \mathrm{Age}$ is the difference between artificial age and best-fitted value. The color of each point presents the density of the samples.} 
\label{fig:reproduceparameters}
\end{figure*}

Two types of parameter error are considered. One is the system error ($\sigma_\mathrm{sys}$) caused by our random spectral models and SED fitting method. The system error is estimated by fitting a new set of artificial SEDs with the same library of $10^5$ spectral models. We produce 5000 new artificial spectra with age, metallicity, and $\tau$ randomly sampled in the same way as described in Section \ref{sec:models}. These new spectra are reddened using the same \citet{car89} extinction law and $A_V$ values are randomly selected from 0 to 3 mag. The IRX of each reddened spectrum is calculated through the energy conservation, assuming that the absorbed energy is reradiated by dust in infrared. Then we apply the same method to constrain the dust extinction by IRX and the same minimization method as shown in Section \ref{sec:fitting} to estimate the stellar population parameters. The standard deviation between the estimated and artificial values for the 5000 spectra is regarded as the system error. The system error for each parameter varies in different regions of the parameter range, so it is calculated at an interval of 0.1 dex for age and metallicity or 0.1 mag for $A_V$ (especially for age; see Figure \ref{fig:reproduceparameters} (d)). Note that the delayed-exponential SFH as adopted in our paper cannot describe the bursts occurring on the top of a continuum SFH \citep{gallazzi09,zibetti09}. This limitation artificially reduces the systematic uncertainty on the stellar population parameters.

The other is the random error ($\sigma_\mathrm{ram}$) caused by the photometric uncertainty, which including both errors of photometry and calibration. The photometric error  comes from the flux statistic noise and fluctuation of the sky background. The calibration error is listed in Table \ref{tab:obs}.  The random error of best-fitted parameters is determined by Monte Carlo simulations: (1) for each SED, we generate 100 randomly perturbed SEDs by adding random Gaussian noises to photometric magnitudes assuming that the photometric uncertainties are from normal distributions; (2) best-fitted parameters of these perturbed SEDs are derived through our $\chi^2$ minimization and constraining method of dust extinction; (3) the standard deviation of the 100 fitted values for each parameter is taken as the random error. The final error of each parameter ($\sigma$) is the combination of both system and random errors, which is expressed as $\sigma=\sqrt{\sigma_\mathrm{sys}^2+\sigma_\mathrm{ram}^2}$.

It is well known that age, metallicity, and dust can be degenerate in their effect on broad-band photometric SED. In this work, the IRX is used to constrain the dust extinction. In addition, the broad bands ranging from ultraviolet to near-infrared and intermediate optical bands help to degrade the age-metallicity degeneracy. Actually, the BATC intermediate bands are more sensitive to the metallicity. As mentioned in \citet{kong00}, the BATC near-infrared color was used to estimate the metallicity when they performed stellar population synthesis fitting. 

We verify whether our method can reliably recover the parameters through Monte-Carlo simulations. A total of 5,000 artificial SEDs are created in the same process as described in Section \ref{sec:models}. These SEDs are scaled to match the mean observed BATC $j$-band magnitude. Random Gaussian noises are then added to these artificial magnitudes according to a given S/N. IRX is simulated through the energy conservation as mentioned in the above error estimation. The median S/N is about  20. Figure \ref{fig:reproduceparameters} gives a comparison between the estimated parameters derived by our SED-fitting method and artificial parameters at a typical S/N of 20. The parameter RMSs for log(Age), log(Z), and $A_V$ at S/N of 20 are about 0.14 dex, 0.07 dex and 0.06 mag, respectively. We also perform Monte-Carlo simulations for different S/Ns to present the RMSs of the best-fit parameters as listed in Table \ref{tab:sigma}. From Figure \ref{fig:reproduceparameters} (a)--(c), we can see that age, metallicity, and dust extinction are well recovered. As shown in Figure \ref{fig:reproduceparameters} (d) and (e), the age scatter becomes larger as it gets older, while the scatter for metallicity does not change with metallicity. Figure \ref{fig:reproduceparameters} (f) shows the difference of metallicity between the artificial and best-fitted values as a function of age and Figure \ref{fig:reproduceparameters} (g) shows the difference of age between the artificial and best-fitted values as a function of metallicity. From these two plots,  there is no obvious bias variation age or metallicity, which indicates that the degeneracy between age and metallicity is not significant.  

\begin{table}
\centering
\caption{Parameter uncertainties estimated by simulations with different S/Ns}
\label{tab:sigma}
\begin{tabular}{cccc}
\hline
S/N & $\sigma$(log age) & $\sigma$(log Z) & $\sigma$($A_{V}$)  \\
      &      (log yr)          &        (dex)           &         (mag)              \\
\hline
10 & 0.180 & 0.097 & 0.070  \\
20 & 0.136 & 0.066 & 0.059  \\
30 & 0.121 & 0.057 & 0.053  \\
50 & 0,104 & 0.046 & 0.050  \\
100 &0.092 & 0.042 & 0.042  \\
\hline
\end{tabular}
\end{table}

\subsection{Fitting examples} \label{subsec:fitting results}
\begin{figure}
\centering
\includegraphics[width=0.5\textwidth,angle=-90]{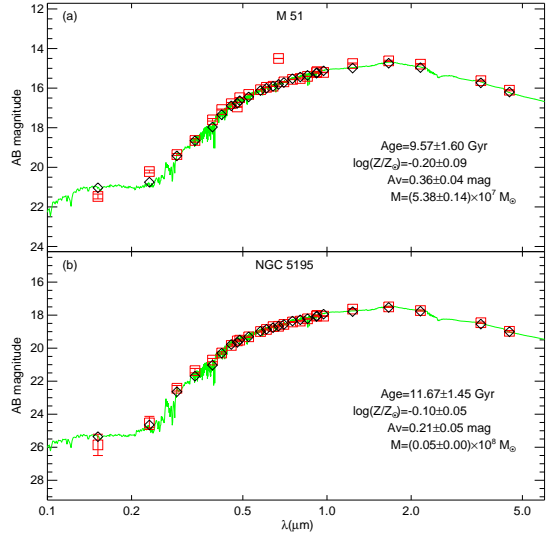}
\caption{Two examples of our stellar population synthesis fitting for several selected SEDs. The SEDs are located at the centers of M 51 (a) and NGC 5195 (b). The red open squares are observed AB magnitudes with vertical bars indicating the photometric errors. The best-matched model magnitudes are plotted in black diamond, and corresponding model spectra are drawn in green curves. The fitted age, metallicity, estimated extinction ($A_V$) by IRX, and inferred stellar mass ($M$) are also displayed in each panel.}
\label{fig:example}
\end{figure}

We choose two pixels at the centers of M 51 and NGC 5195 to demonstrate the observed SEDs, fitted parameters, and best-fitted templates as shown in Figure \ref{fig:example}. The IRAC 3.6 and 4.5 $\mu$m emissions are dominated by starlight, while the 5.6 $\mu$m and 8.0 $\mu$m are mixed by the emission from polycyclic aromatic hydrocarbons (PAHs). Thus, only IRAC 3.6 and 4.5 $\mu$m are used for stellar population analysis. The BATC $a$ band has a low image quality, so we do not use this band.  The BATC $e$ and $i$ bands are also not used due to the strong gaseous emission lines from [O III] and H$\alpha$. Finally, there are a total of 23 bands actually used for fitting. At least 15 bands are required to have the flux S/N larger than 5. Although not used in fitting, the observed points of BATC $a,e,i$ are still overplotted in Figure \ref{fig:example}. In general, our minimization method provides a good match between the model spectrum and observed SED.

\section{Two-dimensional and radial distributions of parameters} \label{sec:result}
By fitting the observed SED with stellar population synthesis models, we can derive a series of parameters including age, metallicity, dust extinction, and stellar mass for each pixel. We extract the multi-band SED pixel by pixel and generate corresponding parameter maps. The pixel scale is about 1\farcs7, which is corresponding to 69 pc at the distance of 8.4 Mpc. It should be noted that all images have been convolved to a PSF FWHM of $\sim$6{\arcsec}, so the actual spatial resolution are larger than 69 pc. After visually examining the parameter maps and multi-band images, we take decl. = 47\fdg24 as the rough boundary of M 51 and NGC 5195.

\subsection{Age}
\begin{figure*}
\centering
\includegraphics[width=0.33\textwidth,angle=-90]{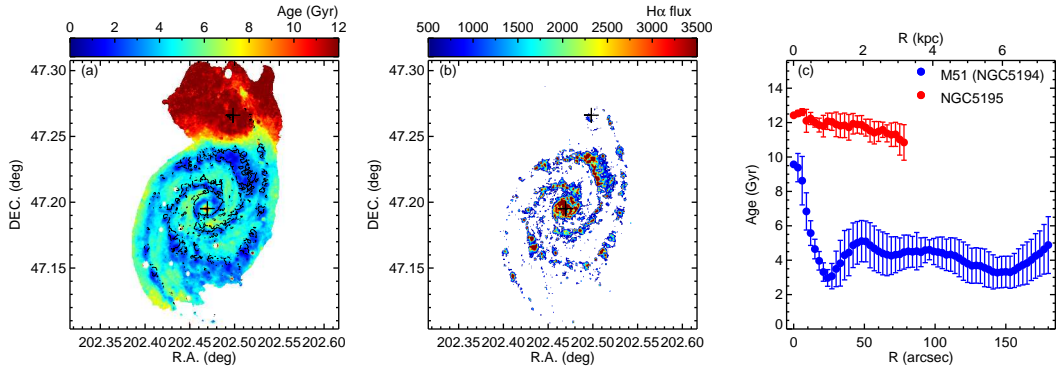}
\caption{Left: age map. The overplotted contours highlighting spiral arms and {\HII} regions are the {\Ha} isophote. Middle: the {\Ha} intensity map obtained from NASA/IPAC Extragalactic Database. The centers of M 51 and NGC 5195 are displayed with pluses. Right: deprojected radial mass-weighted age profiles. The error bars are the standard deviations at different galactocentric annuli. }
\label{fig:age}
\end{figure*}

The age map in the left panel of Figure \ref{fig:age} shows that M 51 is evidently younger than NGC 5195. The average age of M 51 is 4.75 $\pm$ 1.28 Gyr, while that of NGC 5195 is 12.19 $\pm$ 1.16 Gyr. If not explicitly specified, the above average values and those in the following of this paper are calculated as mass-weighted ones. In the map of Figure \ref{fig:age}, we can see that two young spiral arms extend from the center of M 51. The inter-arm regions are relatively older. In order to check wether our age map can reveal the young {\HII} regions, we overlaid the contours of {\Ha} flux density in the left panel of Figure \ref{fig:age}. It can be found that all {\HII} regions with strong {\Ha} emission have young ages.  However, NGC 5195 lacks such young structures as also indicated in the {\Ha} map. The whole galaxy of NGC 5195 is old.

Radial profiles of stellar population parameters are calculated for each galaxy. The two galaxies are deprojected according to their inclination angles of 20{\degr} and 43{\degr} and position angles of 10{\degr} and 91{\degr} for M 51 and NGC 5195 \citep{tully74,spillar92}, respectively. The radial profile is then obtained by computing azimuthal mass-weighted averages in a set of galactocentric annuli. The radial age profile is presented in the right panel of Figure \ref{fig:age}. The center is as old as the surrounding bulge. The bulge of M 51 is small, whose size is about $11\arcsec\times16\arcsec$\citep{lamers02}.  The age decreases steeply from the bulge to the inner disk (about 3 Gyr) at a galactocentric distance of about 25\arcsec (1 kpc).  The age gradient in the outer disk with $r > 25\arcsec$ is quite flat. Such kinds of two different age gradients in the disk are also found in M 33, NGC 628, and M 101 \citep{williams09,zou2011,lin13}. The overall age of NGC 5195 is around 11.5 Gyr and we can also seen that there is a mild radial gradient.

\subsection{Metallicity}
\begin{figure*}
\centering
\includegraphics[width=0.45\textwidth,angle=-90]{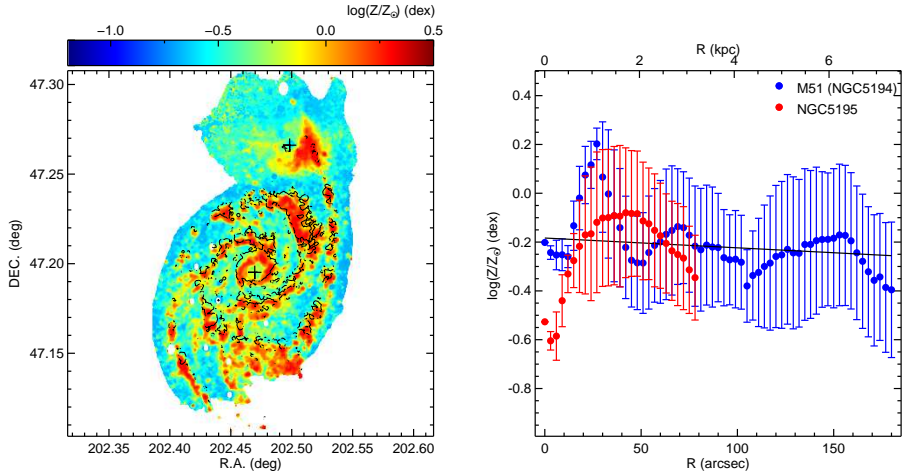}
\caption{Left: metallicity map in $\log(Z/Z_{\sun})$.  The overplotted contours indicate the {\Ha} intensity map. The centers of M 51 and NGC 5195 are displayed with pluses. Right: deprojected radial mass-weighted metallicity profiles. The error bars are the standard deviations at different galactocentric annuli. The black solid line shows the best-linear fit in M 51 using the radial mass-weighted metallicity profiles.}
\label{fig:metallicity}
\end{figure*}

Figure \ref{fig:metallicity} shows the stellar metallicity distributions. The overall average metallicity of M 51 is -0.23 $\pm$ 0.05 dex, while the average abundance of NGC 5195 is about -0.34 $\pm$ 0.05 dex. The metallicity map in Figure \ref{fig:metallicity} presents that many regions in M 51 are super solar. Most of them are related to spiral arms as shown by the contours of the overlaid {\Ha} intensity map, where massive stars were usually born. The average metallicity $\log(Z/Z_{\sun})$ is 0.05 $\pm$ 0.18 dex for these regions. Combining the age and metallicity maps, we can see that the two spiral arms are constituted by young and super-solar stellar populations. Taking the solar abundance of 12 + log(O/H)$_\odot = 8.69$, \citet{croxall15} also estimated the gas-phase metallicity to be solar or slightly super-solar using spectra of multiple {\HII} regions in M51. One of spiral arms spreads into the east of NGC 5195, standing out of the overall poor abundance in this galaxy. The tail of this arm seems to keep the rich chemical composition from the host galaxy of M 51, but have a similar old age as NGC 5195.

The right panel of Figure \ref{fig:metallicity} shows the radial stellar metallicity profile. The disk presents a metallicity gradient of about $\rm -0.016\pm0.014 ~dex~kpc^{-1}$.  There are several studies about gas-phase metallicity with spectroscopy of {\HII} regions in M 51 \citep{zaritshy94,bre04,moustakas10,croxall15}. A weak abundance gradient of $\rm -0.025~\pm~0.008~dex~kpc^{-1}$ was derived by using the auroral lines of 10 {\HII} regions \citep{bre04}. \citet{moustakas10} obtained two gradient values of -0.038 $\pm$ 0.004 dex kpc$^{-1}$ and -0.024 $\pm$ 0.004 dex kpc$^{-1}$. These gradients are calculated by using different metallicity calibration methods. The radial metallicity profile of NGC 5195 shows a peak around $r = 50\arcsec$, mainly caused by the metal-rich spiral arm originating from M 51. 

\subsection{Dust extinction}
\begin{figure*}
\centering
\includegraphics[width=0.33\textwidth,angle=-90]{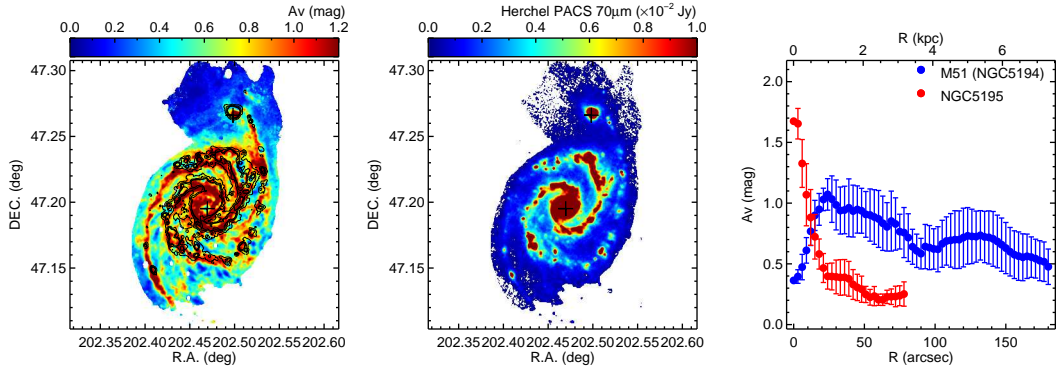}
\caption{Left: dust extinction map in $A_V$. The overplotted contours show the regions with strong $Herschel$ PACS 70 $\mu$m radiation. The centers of M 51 and NGC 5195 are displayed with pluses. Middle: intensity map of $Herschel$ PACS 70 $\mu$m in Jy sr$^{-1}$. Right: deprojected radial mass-weighted extinction profiles. The error bars are the standard deviations at different galactocentric annuli.}
\label{fig:dust}
\end{figure*}

The left panel of Figure \ref{fig:dust} shows the dust extinction map. \citet{looze14} derived the FUV attenuation map (see Figure 6 in their paper) with a resolution of about $12.1\arcsec$ based on FUV and mid- and far-infrared observations. If considering $A_\mathrm{FUV}/A_V \sim 3$ \citep{fis05}, our dust extinction map is quite close to theirs. The average extinction $A_V$ of M 51 is 0.70 $\pm$ 0.04 mag. It is a little high to typical Sbc galaxies with $A_V\sim0.5$ mag \citep{bos03,munoz09}. We overlay the contours of the $Herchel$ PACS 70 $\mu$m intensity map as shown in the middle of Figure \ref{fig:dust} onto the dust extinction map. It can be seen that most of the $70 \mu$m observational features coincide with the substructures with high dust extinction. Although the outer spiral arms present low infrared luminosity, the dust extinction is still high due to resided dust lanes and fewer illuminating young stars. Most of the regions with high extinction are located at the inner part of spiral arms.  

The radial extinction profiles are shown in the right panel of Figure \ref {fig:dust}. The dust attenuation in the bulge of M51 is smaller than the surrounding inner disk. It increases steadily from the center and peaks at $r = 25\arcsec$ (1.0 kpc), forming a young, metal-rich, and dusty ring. The outer disk presents a descending gradient with $A_V$ varying from 1.0 to 0.5 mag. The average dust attenuation of NGC 5195 is 0.65 $\pm$ 0.04 mag. In this galaxy, the central region is dusty, while the outer part is almost free of dust. The nucleus of NGC 5195 has a maximum extinction up to about 1.67 mag. \citet{spillar92} has estimated the extinction in the center to be as high as $A_V=2.0$ mag. Despite lack of star formation, for NGC 5195, it has a relatively high (29 $\pm$ 3 K) dust temperature in the nuclear region, which are associated with actively star forming galaxy under normal circumstances \citep{men12}.  Although there is no obvious recent star formation as shown in the $\Ha$ intensity map, this kind of dusty and relatively hot center can be heated by the strong radiation field from high stellar densities \citep{engelbracht10}. In addition, the tip of the overlaid spiral arm from M 51 is almost dust-free with $A_V \sim 0.2$ mag. There might be some mechanisms leading to the dust loss.

\subsection{Stellar mass surface density}
\begin{figure*}
\centering
\includegraphics[width=0.45\textwidth,angle=-90]{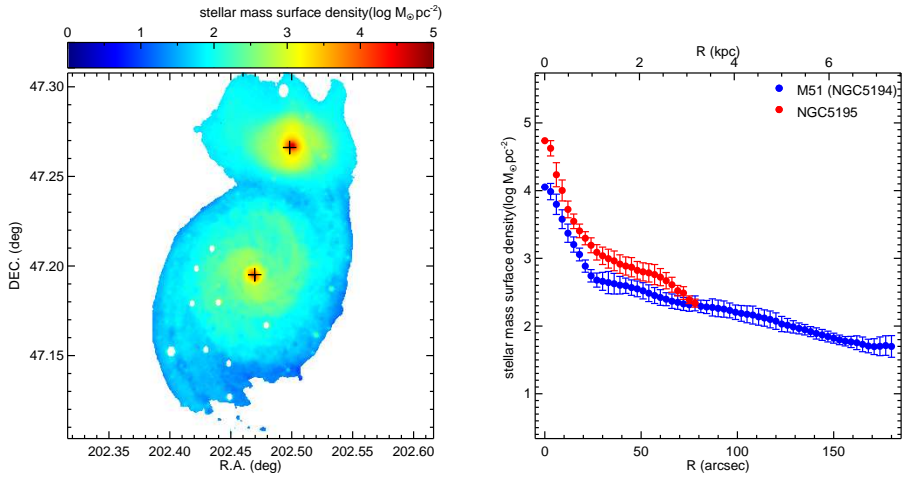}
\caption{Left: stellar mass surface density map. The centers of M 51 and NGC 5195 are displayed with pluses. Right: deprojected radial stellar mass surface density profiles. The error bars are the standard deviations at different galactocentric annuli.}
\label{fig:mass}
\end{figure*}

The map of stellar mass surface density as shown in the left of Figure \ref{fig:mass} follows the $K_s$-band morphology, which is regarded as a good tracer of stellar mass. The bulge is dominated by low-mass stars still on the main sequence contributing most of the total stellar mass. The mass is enhanced along the spiral arms in M 51. The stellar mass surface density radially decreases from the central regions of both galaxies to the outskirts (see the right of Figure \ref{fig:mass}). The radial profiles display distinct bulge and disk components. The core region of NGC 5195 is denser than M 51. The total stellar mass of M 51 is $2.97\pm0.23\times10^{10} M_{\sun}$ and that of NGC 5195 is $2.90\pm0.11\times10^{10} M_{\sun}$. The mass ratio is close to 1:1. The stellar mass of NGC 5195 might be overestimated, because one of the spiral arms extending from M 51 overlays this galaxy. Table \ref{tab:pars} lists the average parameter values for both M 51 and NGC 5195. 

\begin{table*}
\centering
\caption{Average parameter values for M 51 and NGC 5195}
\label{tab:pars}
\begin{tabular}{ccccc}
\hline
& Age (Gyr) & $\log (Z/Z_{\sun}$) (dex) & $A_V$ (mag) & stellar mass$^a$ ($M_{\sun}$) \\
\hline
M 51 & $4.74\pm1.28$ & $-0.23\pm0.05$ & $0.70\pm0.04$  & $2.97\pm0.23\times10^{10}$ \\
\hline
NGC 5195 & $12.19\pm1.16$ & $-0.34\pm0.05$ & $0.65\pm0.04$ & $2.90\pm0.11\times10^{10}$ \\
\hline
\multicolumn{5}{l}{$^a$ Stellar mass is calculated by summing the masses of all pixels belonging to the galaxy.} \\
\end{tabular}
\end{table*}

\subsection{Comparing with the work of \citet{men12}} \label{subsubsec:comparison}

For M51-NGC 5195 system, \citet{men12} also studied the spatially-resolved properties of this system from the multi-wavelength photometric data. In their work, the stellar population and dust properties were independently determined using different bands. The optical and near-infrared bands were used to modeling the stellar population and mid- and far-infrared bands are used to obtain the dust parameters. For the stellar population synthesis, they adopted a sum of a continuous population modeled as an exponentially decreasing function of time and an additional burst of star formation modeled as a tau-exponential with timescale of $\tau$ = 100 Myr. The main goal is to determine when the recent bust of star formation occurred. They indeed obtained a starburst with age of about 370--480 Myr ago, consistent with the simulations. We adopt a single exponentially delayed SFH and give average age of $\sim$ 2--3 Gyr for the {\HII} regions. The age is older than the time of the recent encounter, because the observed SED includes the underlying older disk stellar populations and thus the derived age is the combination of the recently bursted stars and old disc stars. 

However, the issue of parameter degeneracy should be more serious in \citet{men12}, because their stellar population modeling was based on only 7 bands from optical $B$ to 2MASS $K_s$ bands and they considered two components of star formation histories simultaneously. In our work, we gather a total of 26 bands with wavelength ranging from 1500 {\AA} to 4.5 $\mu$m. The BATC intermediate colors can be used to well constrain the metallicity. The observed IRX is adopted to constrain the dust extinction. In this way, the parameter degeneracy is considerably relieved. In addition,  the spatial resolution of our data is about 6\arcsec, about 5 times better than the one in \citet{men12}, so that we can examine the parameter distributions in more details. For example, \citet{men12} gave a flat age profile for both M51 and NGC 5195 ($\sim$ 7 - 10 Gyr). In our age map, we can see an older bulge and younger disk and the age gradient is steeper in the inner region and becomes shallower in the outer region, which is common in many spiral galaxies. Due to the lack of UV data, \citet{men12} claimed that the dust extinction was modestly underestimated ranging from 0.5 mag in the center to 0.2 mag in the outer region. Our dust extinction distribution presents a decreasing gradient ranging from 1.0 to 0.5 mag, which is consistent with the results of \citet{looze14}. Note that the extent of the radial profiles as computed in our work is $\sim$ 7.5 kpc in terms of distance of 8.4 Mpc, smaller than that of \citet{men12} ($\sim$ 12 kpc).

\section{Discussion} \label{sec:discussion}
\subsection{Pseudobulge of M 51 and Secular Evolution} \label{subsec:Pseudobulge}
Resent observations show that bulges in nearby disk galaxies are complicated and there are at least two types: classical bulges and pseudobulges. Classical bulges that are dominated by random motions and old stars. Pseudobulges are characterized by rotating motion, active star formation, nuclear bar, ring and/or spiral,  and near-exponential surface brightness profile \citep{Kormendy04}. 

The S\'{e}rsic index of the M 51's bulge is about $0.55\pm0.07$ \citep{fis10}. It is less than 2.0, which is commonly used as a separation between classical bulges and pseudobulges. From high-resolution images and our parameter maps, we can see resolved nuclear spirals. In addition, it is reported that there is a small bar in the bulge. \citet{menendez-delemestre07} measured the bar fraction and found that the bulge of M 51 has an inner nuclear bar with a major axis length of about 16{\arcsec} and an orientation of about $139^{\circ}$. From our radial parameter profiles as shown in Section \ref{sec:result}, the bulge is metal-poorer and less dusty than the surrounding area. There is a clear $\rm H\alpha$ emission in BATC $i$ band as shown in the panel (a) of Figure \ref{fig:example}, implying some extent of star formation, although a part of the $\rm H\alpha$ emission is contributed by AGN \citep{goad79,moustakas10}. Gas inflow could dilute the metallicity and induce more star formation.  A rather chaotic distribution of dust lanes can be found in high-resolution images of the central region of M 51 taken by the $Hubble ~Space ~Telescope$ (HST), which also indicates that M 51 is transporting gas to the nucleus \citep{grillmair97}. The nucleus of M 51 has an average rotation velocity of about $\rm 240\pm20 ~km ~s^{-1}$ \citep{Kormendy10}, and the velocity dispersion is about $\rm 96.0\pm 8.7~ km ~s^{-1}$ \citep{ho09}. \citet{fis10} measured the $\rm 3.6-8.0\mu m$ color as a rough estimate of the specific SFR and reported that the bulge of M 51 is slightly active. A group of 30 bright massive stars found in HST images also imply that there is ongoing massive star formation \citep{lamers02}. All above characteristics hint that the bulge of M 51 is a slightly active pseudobulge. 

Unlike classical bulges that are typically merger-built, galaxies form pseudobulges via processes in the secular evolution of the galactic disk \citep{Kormendy04,fis08,fisher09,fis10}. The growth of a pseudobulge is driven by the nonaxisymmetry of bars, ovals and/or spiral structures in the disk that can cause gas infall, build up a central mass concentration, and hence trigger the star formation. It is possible that the pseudobulge of M 51 was formed by the secular evolution of the disk driven by the nonaxisymmetric potential of spiral arms and small bar. Since M 51 and NGC 5195 are currently interacting, extremely gas-rich accretion events, gravitational encounters and interactions also have effect on forming the pseudobulge\citep{Kormendy04,fis08}.

\subsection{Evolution effect of the galaxy interaction} \label{subsec:interaction}
The M 51--NGC 5195 system is an interacting galaxy pair. They underwent an encounter 300-500 Myr ago, which is inferred from kinematic and hydrodynamic modelings \citep{salo00,dob10}. There are two grand-design spiral arms in the disk of M 51. The parameter maps in our paper show that these two spectacular spiral arms starting from the bulge are dominated by younger, metal-richer, and less dusty stellar populations. The spiral arms show considerable {\HII} regions, supporting that there is ongoing star formation across the galaxy. It was stated in \citet{nikola01} that the outlying spiral substructures are attributed to material clumping formed by the galactic interaction.  The radial age profile of M 51 presents a flat gradient in the outer disk. The abundance gradient of the stellar content is about $-0.016$ dex kpc$^{-1}$, close to the gas-phase metallicity gradient. It is much lower than other typical isolated galaxies \citep{zaritshy94,lin13}. \citet{salo00} implied that there might be multiple encounters between M 51 and NGC 5195. It is possible that those multiple close encounters make the age and metallicity gradients flatten \citep{kewley06}.

Further, as seen from the age, metallicity, and $A_V$ maps, there are more substructures in the north spiral arms filled with young, metal-rich, and dusty stellar populations. They are also presented in the {\Ha} intensity map. \citet{Kaleida2010} identified and measured 120 single-aged stellar associations. They found that there is an enhancement in the number of stellar associations in the northern spiral arm of M 51. The enhanced star formation in the north arms is possibly induced by the galactic interaction. 

NGC 5195 is a post-starburst galaxy. The old age and a lack of H$\alpha$ emission indicate that there is no recent star formation. It is interesting to find that the bulge of NGC 5195 is fairly dusty in Figure \ref{fig:dust}.  The dust is most likely to be heated by the evolved stellar populations, since the stellar mass is highly concentrated in the galaxy core as shown in the stellar mass distributions of Figure \ref{fig:mass}. It is exciting to find that the tip of the north-east spiral arm extending from M 51 to NGC 5195 is almost free of dust but still keep the high metallicity of the original arm (see maps in Figure \ref{fig:metallicity} and \ref{fig:dust}). By contract, the other spiral arms in the opposite direction present both high metallicity and large dust attenuation. It is possible that the close encounters of these two galaxies make the dust in the north-east arm be accreted by NGC 5195. As analyzed in \citet{men12}, when the two galaxies were at the stages of close encounters, tidal forces likely leaded to high accretion rates of gas and dust that cause starburst. But as the galaxies moved apart, the tidal forces were reduced and gas was exhausted so that the star formation ceased in NGC 5195.  It is also possible that the growth of the stellar bulge of NGC  5195 resulting in a stable disk cause the cease of star formation \citep[termed as ``morphological quenching",][]{martig09}, as we can see that most of the stellar mass and dust are concentrated in the bulge.

In conclusion, there are a lot of hints indicating that the bulge of M51 is a pseudobulge formed through the secular evolution. The gravitational asymmetry drives the gas inflow and leads to the slow growth of the pseudobulge. The galaxy interaction makes more enhanced star formation in the north arms and yields possible gas inflow and material exchange between M51 and NGC 5195. Both secular evolution and galaxy interaction might play important roles on the current evolution of this interacting system.

\section{Summary} \label{sec:summary}
Nearby galaxies such as the M 51--NGC 5195 galaxy pair are ideal objects to examine the stellar population, galaxy interaction, and galaxy evolution in great details. Multi-wavelength photometric data ranging from UV to infrared provide spatially resolved SEDs for nearby galaxies. Through fitting the observed SED with evolutionary stellar population synthesis models, we can reliably derive a series of stellar population properties. Although similar work of \citet{men12} was carried out to present spatially-resolved stellar population analysis for this interacting system, the stellar population modeling of their work is based on only a few bands and two components of SFHs. There are serious parameter degeneracies in \citet{men12}. In this paper, we revisit the stellar population synthesis analysis with a higher spatial resolution and much more bands ranging from UV and near-infrared in order to derive more reliable parameters.

We collect a total of 28 deep images of the M 51--NGC 5195 system from GALEX, XMM-OM, BATC, BASS, MzLS, 2MASS, and Spitzer with wavelength ranging from 1500{\AA} to 24 $\mu$m and a lowest spatial resolution of 6\arcsec. These data are processed by a dedicated pipeline, which includes the process of subtracting sky background, matching to a common resolution of 6{\arcsec} and a pixel scale of 1{\farcs}7, and masking foreground stars. The stellar population synthesis models of BC03 with the \citet{cha03} IMF, Padova 1994 evolutionary tracks, and delayed-exponential SFH are adopted in this paper. From SED fitting, we obtained the two-dimensional distribution and radial profiles of a series of parameters including age, metallicity, dust extinction, and stellar mass. The dust extinction is constrained by the observed IRX.

The average mass-weighted age, metallicity in $\log(Z/Z_{\sun})$, and dust extinction in $A_V$ for M 51 is 4.74$\pm1.28$ Gyr, $-0.23\pm0.05$ dex, and $0.70\pm0.04$ mag, respectively. Those for NGC 5195 are $12.19\pm1.16$ Gyr, $-0.34\pm0.05$ dex, and $0.65\pm0.04$ mag. NGC 5195 is much older. The stellar masses are $2.97\pm0.23\times10^{10}$ $M_{\sun}$ for M 51 and  $2.90\pm0.11\times10^{10}$ $M_{\sun}$ for NGC 5195. The mass ratio is close to 1:1.  The grand-design spiral arms can be clearly seen in almost all parameter maps. Generally, they are young, metal-rich, dusty, and massive. The enhanced star formation in the north arms is possibly induced by the galaxy interaction. However, except for the spiral arm extending from M 51 into NGC 5195, the stellar population properties of NGC 5195 are quite featureless. There are very small radial age and metallicity gradients in the outer disk of M 51. The metallicity gradient is about -0.016 dex kpc$^{-1}$, close to gas-phase one. It is possible that the galaxy interaction between M 51 and NGC 5195 makes the gradients more flatten than other isolated nearby galaxies. The core region of NGC 5195 is dusty and has exceptional high dust temperatures. The dust content might be accreted from M 51 through close encounters, since the tip of the superposed spiral arm from M 51 seems free of dust. The large amount of old stellar populations in the core of NGC 5195 can heat the dust. The secular evolution and galaxy interaction might jointly lead the current evolution and structure growth of the M 51--NGC 5195 system. 

\begin{acknowledgements}
We thank the anonymous referee for his/her thoughtful comments and insightful suggestions that improve our paper greatly. This work is supported by Major Program of National Natural Science Foundation of China (No. 11890691). This work is also supported by the National Basic Research Program of China (973 Program; grant Nos.\ 2017YFA0402600), the National Natural Science Foundation of China (NSFC; grant Nos.\ 11433005, 11673027, 11733007, 11973038, 11320101002, and 11421303), and the External Cooperation Program of Chinese Academy of Sciences (grant No.\ 114A11KYSB20160057). The 60/90 cm Schmidt telescope at the Xinglong Station of National Astronomical Observatory of China is jointly operated and administrated by the National Astronomical Observatories of China and Center for Astronomical Mega-Science, Chinese Academy of Sciences.

This work is based in part on observations made with the $Spitzer ~Space ~Telescope$, which is operated by the Jet Propulsion Laboratory, California Institute of Technology under a contract with NASA. This publication make use of data products from the Two Micron All Sky Survey, which is a joint project of the University of Massachusetts and the Infrared Processing and Analysis Center/California Institute of Technology, funded by the National Aeronautics and Space Administration and the National Science Foundation.

The BASS is a collaborative program between the National Astronomical Observatories of Chinese Academy of Science and Steward Observatory of the University of Arizona. It is a key project of the Telescope Access Program (TAP), which has been funded by the National Astronomical Observatories of China, the Chinese Academy of Sciences (the Strategic Priority Research Program ``The Emergence of Cosmological Structures" grant No. XDB09000000), and the Special Fund for Astronomy from the Ministry of Finance. The BASS is also supported by the External Cooperation Program of Chinese Academy of Sciences (Grant No. 114A11KYSB20160057) and Chinese National Natural Science Foundation (Grant No. 11433005). 
\end{acknowledgements}


\begin{thebibliography}{99}

  \bibitem[Bertin(2010)]{bertin02} Bertin, E.\ 2010, SWarp: Resampling and Co-adding FITS Images Together, ascl:1010.068

  \bibitem[Boselli et al.(2003)]{bos03} Boselli, A., Gavazzi, G., \& Sanvito, G.\ 2003, \aap, 402, 37
  
  \bibitem[Bresolin et al.(2004)]{bre04} Bresolin, F., Garnett, D.~R., \& Kennicutt, R.~C., Jr.\ 2004, \apj, 615, 228

  \bibitem[Bruzual A.~\& Charlot(1993)]{bru93} Bruzual A., G., \& Charlot, S.\ 1993, \apj, 405, 538

  \bibitem[Bruzual \& Charlot(2003)]{bru03} Bruzual, G., \& Charlot, S.\ 2003, \mnras, 344, 1000

  \bibitem[Bruzual(2007)]{bru07} Bruzual, A.~G.\ 2007, Stellar Populations as Building Blocks of Galaxies, 241, 125
  
  \bibitem[Burstein et al.(1994)]{bur94} Burstein, D., Hester, J.~J., Windhorst, R.~A., et al.\ 1994, Bulletin of the American Astronomical Society, 26, 41.10
  
  \bibitem[Calzetti et al.(2005)]{cal05} Calzetti, D., Kennicutt, R.~C., Jr., Bianchi, L., et al.\ 2005, \apj, 633, 871
  
  \bibitem[Cardelli et al.(1989)]{car89} Cardelli, J.~A., Clayton, G.~C., \& Mathis, J.~S.\ 1989, \apj, 345, 245
  
  \bibitem[Chabrier(2003)]{cha03} Chabrier, G.\ 2003, \pasp, 115, 763
  
  \bibitem[Chambers et al.(2016)]{cha16} Chambers, K.~C., Magnier, E.~A., Metcalfe, N., et al.\ 2016, arXiv:1612.05560
  
  \bibitem[Charlot \& Bruzual A(1991)]{cha91} Charlot, S., \& Bruzual A, G.\ 1991, \apj, 367, 126
  
  \bibitem[Cort{\'e}s et al.(2006)]{cor06} Cort{\'e}s, J.~R., Kenney, J.~D.~P., \& Hardy, E.\ 2006, \aj, 131, 747
  
  \bibitem[Croxall et al.(2015)]{croxall15} Croxall, K.~V., Pogge, R.~W., Berg, D.~A., et al.\ 2015, \apj, 808, 42.
  
  \bibitem[Dark Energy Survey Collaboration et al.(2016)]{des16} Dark Energy Survey Collaboration, Abbott, T., Abdalla, F.~B., et al.\ 2016, \mnras, 460, 1270
  
  \bibitem[De Looze et al.(2014)]{looze14} De Looze, I., Fritz, J., Baes, M., et al.\ 2014, \aap, 571, A69
  
  \bibitem[Dey et al.(2019)]{dey18} Dey, A., Schlegel, D.~J., Lang, D., et al.\ 2019, \aj, 157, 168
  
  \bibitem[Dobbs et al.(2010)]{dob10} Dobbs, C.~L., Theis, C., Pringle, J.~E., \& Bate, M.~R.\ 2010, \mnras, 403, 625
  
  \bibitem[Eldridge \& Stanway(2009)]{eld09} Eldridge, J.~J., \& Stanway, E.~R.\ 2009, \mnras, 400, 1019
  
  \bibitem[Engelbracht et al.(2010)]{engelbracht10} Engelbracht, C.~W., Hunt, L.~K., Skibba, R.~A., et al.\ 2010, \aap, 518, L56

  
  \bibitem[Fan et al.(1996)]{fan96} Fan, X., Burstein, D., Chen, J.-S., et al.\ 1996, \aj, 112, 628
  
  \bibitem[Feldmeier et al.(1997)]{fel97} Feldmeier, J.~J., Ciardullo, R., \& Jacoby, G.~H.\ 1997, \apj, 479, 231
  
  \bibitem[Fioc \& Rocca-Volmerange(1997)]{fio97} Fioc, M., \& Rocca-Volmerange, B.\ 1997, \aap, 326, 950
  
  \bibitem[Fischera \& Dopita(2005)]{fis05} Fischera, J., \& Dopita, M.\ 2005, \apj, 619, 340
  
  \bibitem[Fisher \& Drory(2008)]{fis08} Fisher, D.~B., \& Drory, N.\ 2008, \aj, 136, 773
  
  \bibitem[Fisher \& Drory(2010)]{fis10} Fisher, D.~B., \& Drory, N.\ 2010, \apj, 716, 942
  
  \bibitem[Fisher et al.(2009)]{fisher09} Fisher, D.~B., Drory, N., \& Fabricius, M.~H.\ 2009, \apj, 697, 630
  
  \bibitem[Gaia Collaboration et al.(2016)]{gai16} Gaia Collaboration, Brown, A.~G.~A., Vallenari, A., et al.\ 2016a, \aap, 595, A2
  
  \bibitem[Gallazzi, \& Bell(2009)]{gallazzi09} Gallazzi, A., \& Bell, E.~F.\ 2009, \apjs, 185, 253
  
  \bibitem[Goad et al.(1979)]{goad79} Goad, J.~W., de Veny, J.~B., \& Goad, L.~E.\ 1979, \apjs, 39, 439
  
  \bibitem[Grillmair et al.(1997)]{grillmair97} Grillmair, C.~J., Faber, S.~M., Lauer, T.~R., et al.\ 1997, \aj, 113, 225
  
  \bibitem[Hao et al.(2011)]{hao11} Hao, C.-N., Kennicutt, R.~C., Johnson, B.~D., et al.\ 2011, \apj, 741, 124
  
  \bibitem[Ho et al.(2009)]{ho09} Ho, L.~C., Greene, J.~E., Filippenko, A.~V., \& Sargent, W.~L.~W.\ 2009, \apjs, 183, 1
  
  \bibitem[Jarrett et al.(2003)]{jarrett03} Jarrett, T.~H., Chester, T., Cutri, R., Schneider, S.~E., \& Huchra, J.~P.\ 2003, \aj, 125, 525
  
  \bibitem[Kaleida \& Scowen(2010)]{Kaleida2010} Kaleida, C., \& Scowen, P.~A.\ 2010, \aj, 140, 379
  
  \bibitem[Kennicutt et al.(2003)]{kennicutt03} Kennicutt, R.~C., Jr., Bresolin, F., \& Garnett, D.~R.\ 2003, \apj, 591, 801
  
  \bibitem[Kewley et al.(2006)]{kewley06} Kewley, L.~J., Geller, M.~J., \& Barton, E.~J.\ 2006, \aj, 131, 2004
  

  
  \bibitem[Kong et al.(2000)]{kong00} Kong, X., Zhou, X., Chen, J., et al.\ 2000, \aj, 119, 2745
  
  \bibitem[Kotulla et al.(2009)]{kotulla09} Kotulla, R., Fritze, U., Weilbacher, P., \& Anders, P.\ 2009, \mnras, 396, 462
  
  \bibitem[Kormendy \& Kennicutt(2004)]{Kormendy04} Kormendy, J., \& Kennicutt, R.~C., Jr.\ 2004, \araa, 42, 603
  
  \bibitem[Kormendy et al.(2010)]{Kormendy10} Kormendy, J., Drory, N., Bender, R., \& Cornell, M.~E.\ 2010, \apj, 723, 54
  
  
  \bibitem[Kuntz et al.(2008)]{kuntz08} Kuntz, K.~D., Harrus, I., McGlynn, T.~A., Mushotzky, R.~F., \& Snowden, S.~L.\ 2008, \pasp, 120, 740
  
  \bibitem[Li et al.(2004a)]{li04a} Li, J., Ma, J., Zhou, X., et al.\ 2004a, \aap, 420, 89
  
  \bibitem[Li et al.(2004b)]{li04b} Li, J.-L., Zhou, X., Ma, J., \& Chen, J.-S.\ 2004b, \cjaa, 4, 143
  
  \bibitem[Lin et al.(2013)]{lin13} Lin, L., Zou, H., Kong, X., et al.\ 2013, \apj, 769, 127
  
  \bibitem[Lamers et al.(2002)]{lamers02} Lamers, H.~J.~G.~L.~M., Panagia, N., Scuderi, S., et al.\ 2002, \apj, 566, 818
  
  \bibitem[Leitherer et al.(1999)]{leitherer99} Leitherer, C., Schaerer, D., Goldader, J.~D., et al.\ 1999, \apjs, 123, 3
  
  \bibitem[Maraston(2005)]{maraston05} Maraston, C.\ 2005, \mnras, 362, 799
  
  \bibitem[Martin et al.(2005)]{martin05} Martin, D.~C., Fanson, J., Schiminovich, D., et al.\ 2005, \apjl, 619, L1
  
  \bibitem[Mason et al.(2001)]{mas01} Mason, K.~O., Breeveld, A., Much, R., et al.\ 2001, \aap, 365, L36
  
  \bibitem[Martig et al.(2009)]{martig09} Martig, M., Bournaud, F., Teyssier, R., et al.\ 2009, \apj, 707, 250
  
  \bibitem[Men{\'e}ndez-Delmestre et al.(2007)]{menendez-delemestre07} Men{\'e}ndez-Delmestre, K., Sheth, K., Schinnerer, E., Jarrett, T.~H., \& Scoville, N.~Z.\ 2007, \apj, 657, 790
  
  \bibitem[Mentuch Cooper et al.(2012)]{men12} Mentuch Cooper, E., Wilson, C.~D., Foyle, K., et al.\ 2012, \apj, 755, 165
  
  \bibitem[Meurer et al.(1999)]{meurer99} Meurer, G.~R., Heckman, T.~M., \& Calzetti, D.\ 1999, \apj, 521, 64
  
  \bibitem[Morrissey et al.(2007)]{morrissey07} Morrissey, P., Conrow, T., Barlow, T.~A., et al.\ 2007, \apjs, 173, 682
  
  \bibitem[Moustakas et al.(2010)]{moustakas10} Moustakas, J., Kennicutt, R.~C., Jr., Tremonti, C.~A., et al.\ 2010, \apjs, 190, 233
  
  \bibitem[Mu{\~n}oz-Mateos et al.(2009)]{munoz09} Mu{\~n}oz-Mateos, J.~C., Gil de Paz, A., Boissier, S., et al.\ 2009, \apj, 701, 1965
  
  \bibitem[Nikola et al.(2001)]{nikola01} Nikola, T., Geis, N., Herrmann, F., et al.\ 2001, \apj, 561, 203
  
  \bibitem[Oke \& Gunn(1983)]{oke83} Oke, J.~B., \& Gunn, J.~E.\ 1983, \apj, 266, 713
  
  \bibitem[Ro{\v s}kar et al.(2008)]{roskar08} Ro{\v s}kar, R., Debattista, V.~P., Stinson, G.~S., et al.\ 2008, \apjl, 675, L65
  
  \bibitem[Salo \& Laurikainen(2000)]{salo00} Salo, H., \& Laurikainen, E.\ 2000, \mnras, 319, 393
  
  \bibitem[Schlegel et al.(1998)]{schlegel98} Schlegel, D.~J., Finkbeiner, D.~P., \& Davis, M.\ 1998, \apj, 500, 525
  
  \bibitem[Silva et al.(2016)]{silva16} Silva, D. J., Blum, R. D., Allen, L., et al.\ 2016, in AAS Meeting \#228, 317.02
  
  \bibitem[Skrutskie et al.(2006)]{skrutskie06} Skrutskie, M.~F., Cutri, R.~M., Stiening, R., et al.\ 2006, \aj, 131, 1163
  
  \bibitem[Spillar et al.(1992)]{spillar92} Spillar, E.~J., Oh, S.~P., Johnson, P.~E., \& Wenz, M.\ 1992, \aj, 103, 793
  
  \bibitem[Toomre \& Toomre(1972)]{toomre72} Toomre, A., \& Toomre, J.\ 1972, \apj, 178, 623
  
  \bibitem[Tully(1974)]{tully74} Tully, R.~B.\ 1974, \apjs, 27, 449
  
  \bibitem[Walter et al.(2008)]{walter08} Walter, F., Brinks, E., de Blok, W.~J.~G., et al.\ 2008, \aj, 136, 2563
  
  \bibitem[Werner et al.(2004)]{Werner04} Werner, M.~W., Roellig, T.~L., Low, F.~J., et al.\ 2004, \apjs, 154, 1
  
  \bibitem[White \& Rees(1978)]{white78} White, S.~D.~M., \& Rees, M.~J.\ 1978, \mnras, 183, 341
  
  \bibitem[Williams et al.(2009)]{williams09} Williams, B.~F., Dalcanton, J.~J., Dolphin, A.~E., Holtzman, J., \& Sarajedini, A.\ 2009, \apjl, 695, L15
  
  \bibitem[York et al.(2000)]{york00} York, D.~G., Adelman, J., Anderson, J.~E., Jr., et al.\ 2000, \aj, 120, 1579
  
  \bibitem[Zacharias et al.(2010)]{zacharias10} Zacharias, N., Finch, C., Girard, T., et al.\ 2010, \aj, 139, 2184
  
  \bibitem[Zaritsky et al.(1994)]{zaritshy94} Zaritsky, D., Kennicutt, R.~C., Jr., \& Huchra, J.~P.\ 1994, \apj, 420, 87
  
  \bibitem[Zhou et al.(2001)]{zhou01} Zhou, X., Jiang, Z.-J., Xue, S.-J., et al.\ 2001, \cjaa, 1, 372
  
  \bibitem[Zibetti et al.(2009)]{zibetti09} Zibetti, S., Charlot, S., \& Rix, H.-W.\ 2009, \mnras, 400, 1181.
  
  \bibitem[Zou et al.(2011)]{zou2011} Zou, H., Zhang, W., Yang, Y., et al.\ 2011, \aj, 142, 16
  
  \bibitem[Zou et al.(2017a)]{zou2017} Zou, H., Zhou, X., Fan, X., et al.\ 2017a, \pasp, 129, 064101
  
  \bibitem[Zou et al.(2017b)]{zou2017aj} Zou, H., Zhang, T., Zhou, Z., et al.\ 2017b, \aj, 153, 276
  
  \bibitem[Zou et al.(2018)]{zou18} Zou, H., Zhang, T., Zhou, Z., et al.\ 2018, \apjs, 237, 37

\end{thebibliography}


\end{document}